# QUANTUM ZERO-KNOWLEDGE PROTOCOL USING QUANTUM BIT COMMITMENT WITHOUT QUANTUM MEMORY

Rubens Viana Ramos and José Cláudio do Nascimento
rubens@deti.ufc.br          claudio@deti.ufc.br
*Departament of Teleinformatic Engineering, Federal University of Ceara - DETI/UFC, C.P. 6007 – Campus do Pici - 60755-640 Fortaleza-CE Brasil*

Zero-knowledge proof system is an important protocol that can be used as a basic block for construction of other more complex cryptographic protocols. Quantum zero-knowledge protocols have been proposed but, since their implementation requires advanced quantum technology devices, experimental implementation of zero-knowledge protocols have not being reported. In this work, we present a quantum zero-knowledge protocol based on a quantum bit commitment protocol that can be implemented with today technology. Hence, our quantum zero-knowledge protocol can be readily implemented.

## 1. Introduction

A zero-knowledge system is a protocol between two partners where one of them is the prover, $P$, and the other is the Verifier, $V$. Basically, $P$ has a secret (for example, he knows how to solve an specific problem) and $V$ wants to be sure about that. Thus, $V$ challenges $P$ (asking him to solve the problem) and $P$ has to take actions that proof to $V$ that he really has the secret without allowing $V$ to discover it.

In [1] it was proved that zero-knowledge protocols for NP languages can be implemented if a bit commitment protocol is available. Here, we make a direct extension proposing a quantum zero-knowledge protocol that employs a quantum bit commitment (QBC) protocol. We are going to use the same example used in [1], the 3-colorable graph problem, that is known to be a NP complete problem. The QBC protocol used is the one proposed in [2] that uses coherent states and does not require quantum memory, hence, it can be implemented with today technology. The QBC setup is shown in Fig. 1 and it can be briefly explained as follows:

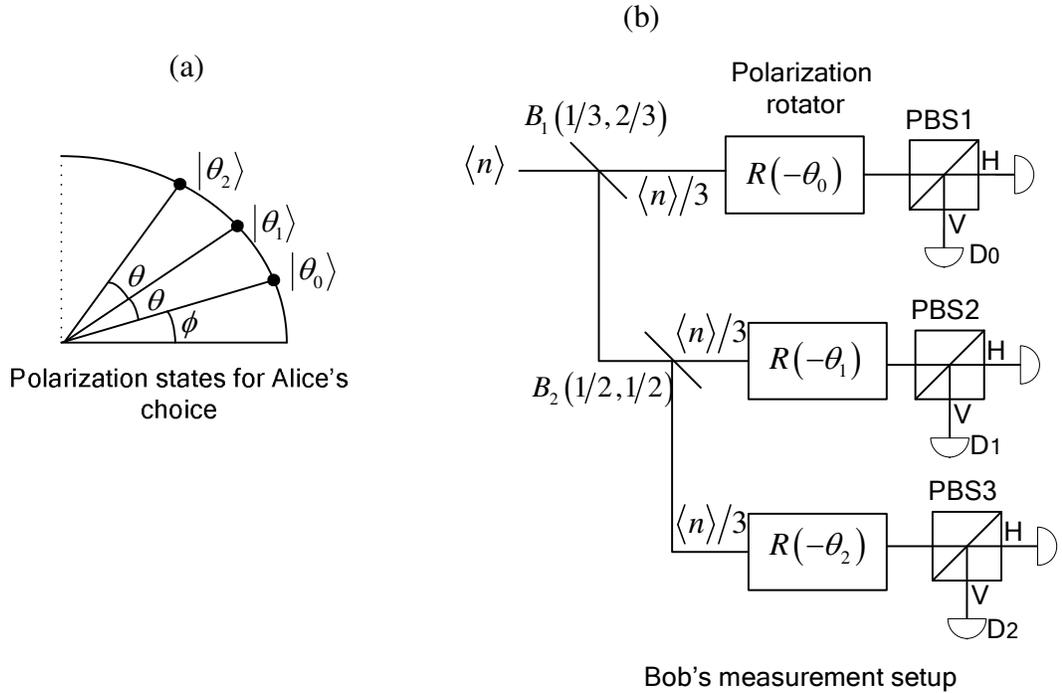

Fig. 1 - Quantum bit/bit string commitment protocol using coherent state. $B_1$ and $B_2$ are beam splitters, PBS is a polarization beam splitter and $R$ is a polarization rotator.

Commit stage:

1. Alice chooses one of the states $\{|\alpha\cos(\phi),\alpha\sin(\phi)\rangle, |\alpha\cos(\phi+\theta),\alpha\sin(\phi+\theta)\rangle, |\alpha\cos(\phi+2\theta),\alpha\sin(\phi+2\theta)\rangle\}$

   ($\langle n\rangle=\alpha^2$) and sends it to Bob.

2. Bob uses the apparatus shown in Fig. 1b to measure the quantum state sent by Alice and he stores the classical results obtained in his ordinary classical memory (e. g. memory of a computer).

Unveil stage:

1. Alice informs to Bob the quantum state sent.

2. Bob checks if Alice's information is in accordance with the results of his measurements. For example, let us suppose Bob had detection in all horizontal outputs and in the vertical output $D_2$. If Alice says to him that she sent $|\theta_2\rangle$ Bob will know she is lying.

Using $\phi=\pi/6.7$ and $\theta=\pi/10$ and $\langle n\rangle=\alpha^2=20$, the probability of Bob cheating Alice is $P_B \approx 0.4$ (probability of Bob to discover the quantum state identity without Alice's information) while the probability of Alice cheating Bob is also $P_A \approx 0.4$ (probability of Alice does not being caught lying during the unveil stage when she informs that sent a quantum state different from the one really sent) [2].

Now, let us turn back to the quantum zero-knowledge protocol. The prover **P** says to verifier **V** that a given graph $G(V,E)$ is 3-colorable, that is, it is possible to label every vertex of $G$ with one of the colors $\{B,R,Y\}$ such that every two adjacent vertices are assigned different colors. In $G(V,E)$, $V$ is the set of vertices $\{1,2,…,n\}$, $|V|=n$ while $E$ is the set of edges, $|E|=m$. If the graph is simple and connected, then $n-1 \leq m \leq n^2/2$. If a graph is 3-colorable then, given a solution, any permutation of the colors is also a solution. Let us call $S_3$ the set of permutations of a colorable graph. The quantum zero-knowledge protocol is described below:

1. **P** chooses randomly one of the permutations in $S_3$. For each vertex, **P** sends a coherent state with $\langle n \rangle =20$ to **V**, where the coherent state sent is chosen according to the following map: vertex $B \to |\alpha\cos(\phi), \alpha\sin(\phi)\rangle$, vertex $R \to |\alpha\cos(\phi+\theta), \alpha\sin(\phi+\theta)\rangle$ and vertex $Y \to |\alpha\cos(\phi+2\theta), \alpha\sin(\phi+2\theta)\rangle$.
2. **V** chooses randomly an edge '$e$' and sends it to **P**.
3. **P** informs to **V** the quantum states sent that correspond to the vertices that belong to the edge '$e$'.
4. According to the QBC protocol, **V** checks if the information sent by **P** really corresponds to the quantum states measured by **V**, if the two quantum states are different and if they belong to the set $\{|\alpha\cos(\phi), \alpha\sin(\phi)\rangle, |\alpha\cos(\phi+\theta), \alpha\sin(\phi+\theta)\rangle, |\alpha\cos(\phi+2\theta), \alpha\sin(\phi+2\theta)\rangle\}$.

These four steps are repeated $m^2$ times and **V** accepts that the graph $G$ is 3-colorable if the step 4 has always a positive checking.

As explained in the QBC protocol, **V** can determine the quantum state sent by **P** without his help with probability little bit lower than 0.4. Thus, the probability of **V** to determine all the colors of the graph alone is $(0.4)^n$. **V** can try this $m^2$ times. On the other hand, if the graph is not colorable, the probability of **P** cheating **V** successfully, per each round, is:

$$\frac{m-1}{m} + \frac{1}{m}\left(\frac{1}{2}0.4 + \frac{1}{2}0.4\right) = \left(1 - \frac{0.6}{m}\right) \qquad (1)$$

Hence, the probability of $P$ cheating $V$ successfully during all the protocol is $(1 - 0.6/m)^{m^2} \approx \exp(-0.6m)$.

## Acknowledgements

This work was supported by the Brazilian agency FUNCAP.